\documentclass[aps,twocolumn,tightenlines,amsmath,amssymb,superscriptaddress]{revtex4-2}

\usepackage{graphicx}
\usepackage{amsmath}
\usepackage{amssymb}
\usepackage{epstopdf}
\usepackage{color}
\usepackage{lipsum}
\usepackage{braket}

\usepackage[colorlinks=true]{hyperref}
\hypersetup{citecolor=blue}

\usepackage{ulem}
\usepackage[dvipsnames]{xcolor}
\usepackage{todonotes}
\DeclareGraphicsExtensions{.eps}

\newcommand{\blue}[1]{\textcolor{blue}{#1}}

\usepackage{tikz}
\begin{document}

\title{Deterministic quantum dot single-photon sources: operational principles and state-of-the-art specifications}

\author{J.~C.~Loredo}
\email{juan.loredo@sparrowquantum.com}
\affiliation{Sparrow Quantum ApS, Nordre Fasanvej 215, 2000 Frederiksberg, Copenhagen, Denmark}

\author{L.~Stefan}
\affiliation{Sparrow Quantum ApS, Nordre Fasanvej 215, 2000 Frederiksberg, Copenhagen, Denmark}

\author{B.~Krogh}
\affiliation{Sparrow Quantum ApS, Nordre Fasanvej 215, 2000 Frederiksberg, Copenhagen, Denmark}

\author{R.~Jensen}
\affiliation{Sparrow Quantum ApS, Nordre Fasanvej 215, 2000 Frederiksberg, Copenhagen, Denmark}

\author{I.~Suleiman}
\affiliation{Sparrow Quantum ApS, Nordre Fasanvej 215, 2000 Frederiksberg, Copenhagen, Denmark}

\author{S.~Kr\"uger}
\affiliation{Lehrstuhl f\"ur Angewandte Festk\"orperphysik, Ruhr-Universit\"at Bochum, Universit\"atsstra{\ss}e 150, 44801 Bochum, Germany}

\author{M.~Bergamin}
\affiliation{Sparrow Quantum ApS, Nordre Fasanvej 215, 2000 Frederiksberg, Copenhagen, Denmark}

\author{H.~Thyrrestrup}
\affiliation{Sparrow Quantum ApS, Nordre Fasanvej 215, 2000 Frederiksberg, Copenhagen, Denmark}

\author{S.~Budtz}
\affiliation{Sparrow Quantum ApS, Nordre Fasanvej 215, 2000 Frederiksberg, Copenhagen, Denmark}

\author{J.~Roulund}
\affiliation{Sparrow Quantum ApS, Nordre Fasanvej 215, 2000 Frederiksberg, Copenhagen, Denmark}

\author{Z.~Liu}
\affiliation{Sparrow Quantum ApS, Nordre Fasanvej 215, 2000 Frederiksberg, Copenhagen, Denmark}

\author{X.~Zhao}
\affiliation{Sparrow Quantum ApS, Nordre Fasanvej 215, 2000 Frederiksberg, Copenhagen, Denmark}

\author{L.~Vertchenko}
\affiliation{Sparrow Quantum ApS, Nordre Fasanvej 215, 2000 Frederiksberg, Copenhagen, Denmark}

\author{A.~Ludwig}
\affiliation{Lehrstuhl f\"ur Angewandte Festk\"orperphysik, Ruhr-Universit\"at Bochum, Universit\"atsstra{\ss}e 150, 44801 Bochum, Germany}

\author{O.~A.D.~Sandberg}
\email{osan@sparrowquantum.com}
\affiliation{Sparrow Quantum ApS, Nordre Fasanvej 215, 2000 Frederiksberg, Copenhagen, Denmark}

\author{P.~Lodahl}
\email{peter.lodahl@sparrowquantum.com}
\affiliation{Sparrow Quantum ApS, Nordre Fasanvej 215, 2000 Frederiksberg, Copenhagen, Denmark}

\begin{abstract}
Non-classical states of light play a fundamental role in quantum technology. From photonic quantum computers and simulators, to quantum communication and sensing, quantum states of light enable performing tasks that may outperform their best classical counterparts. Semiconductor quantum dots embedded in photonic nanostructures offer the most advanced classes of quantum light sources. Importantly, the underlying physics processes determining device performance are today fully understood, and dedicated engineering projects are currently advancing these sources towards real-world quantum technology applications. We review the performance of deterministic single-photon sources based on quantum dots in photonic crystal waveguides, the approach with the highest performance specs since it intrinsically combines suppression of leaky modes and Purcell enhancement to slow-light waveguide mode. Furthermore, we present prototype data from sources that today are commercially available and with performance metrics approaching the ideal.
\end{abstract}

\maketitle

\section{Introduction}
Quantum photonics plays a foundational role in quantum technology. Photons are the sole qubit contender for long-distance quantum communication underpinning the quantum internet~\cite{kimbleQuantumInternet2008} and photonic quantum computing is gaining significant momentum, offering a highly modular and manufacturable route to utility-scale machines~\cite{bartolucciFusionbasedQuantumComputation2023,alexanderManufacturablePlatformPhotonic2025,chan2025practicalblueprintlowdepthphotonic}. The lack of high-efficiency sources of high-quality photonic qubits has been a bottleneck, yet a route to circumvent this deficit has been identified by exploiting massive multiplexing of probabilistic parametric sources~\cite{muxSPS}. While this route is feasible, the corresponding resource requirements for large-scale fault-tolerant quantum computers are daunting and consume an immense amount of energy to operate the many sources, switches, and heralding detectors. 
The advent of deterministic photon sources can potentially disrupt this shortcoming; with solid-state photon source technology maturing in recent years, its transformative promise is starting to materialise.

Quantum dot (QD) sources based on III-V semiconductors have matured to the point that they are now being pursued as a viable commercial technology. This approach exploits a range of advantageous features \cite{lodahlInterfacingSinglePhotons2015}: i) III-V direct bandgap semiconductors allow near-unity internal quantum efficiency, which makes them the material platform of choice also for classical integrated light sources LEDs and lasers~\cite{pimputkarProspectsLEDLighting2009,kramesStatusFutureHighPower2007}. ii) Molecular-beam epitaxy of composite materials allows growth of ultra-clean quantum emitters with controllable composition and optical properties, including tailoring the emission wavelength. These QDs can exhibit collective superradiant enhancement to obtain very high photon rate generation due to large oscillator strengths~\cite{tighineanuSinglePhotonSuperradianceQuantum2016}. iii) QDs can be readily implemented in high-quality nanophotonic cavities and waveguides for realising photonic chips with near-unity coupling efficiency~\cite{arcariNearUnityCouplingEfficiency2014}. iv) scalable QD planar photonic circuits can be realised~\cite{paponIndependentOperationTwo2023} and hetereogeneous integration of QD sources to foundry-fabricated photonic integrated circuits has been developed~\cite{davancoHeterogeneousIntegrationOnchip2017}. v) The short spin-coherence time inherent to III-V materials can be mitigated by nuclear-polarization schemes~\cite{ethier-majcherImprovingSolidStateQubit2017}, which enable the deterministic generation of multi-particle entangled resource states, such as GHZ and cluster states~\cite{lindnerProposalPulsedOnDemand2009,tiurevHighfidelityMultiphotonentangledCluster2022a,mengDeterministicPhotonSource2024}. Recently, a full blueprint for a fault-tolerant photonic quantum computer utilizing deterministic QD sources has been presented~\cite{chan2025practicalblueprintlowdepthphotonic}, promising a highly resource-efficient route to large-scale quantum computing. 

The present manuscript focuses on the performance of deterministic single-photon emission based on QDs in photonic chips. Importantly, the underlying physical principles determining the source performance have been fully established~\cite{lodahlInterfacingSinglePhotons2015}, including the role of enhancing the light-matter coupling in photonic waveguides and cavities, and the identification of the relevant decoherence processes. As a consequence, QD-based deterministic single-photon sources have reached a level of maturity where they are ready for commercialisation and manufacturing. We present data from state-of-the-art planar QD single-photon sources and benchmark their performance compared to alternative approaches. We focus here on sources emitting in the near-infrared at around 930 nm, but recent breakthroughs have shown that similar device performance can be realised in the technologically important telecom O-band~\cite{albrechtsen2025quantumcoherentphotonemitterinterfaceoriginal}.

\section{Operational principle of planar deterministic single-photon sources}

\begin{figure}[htb!]
\centering
\includegraphics[width=0.95\linewidth]{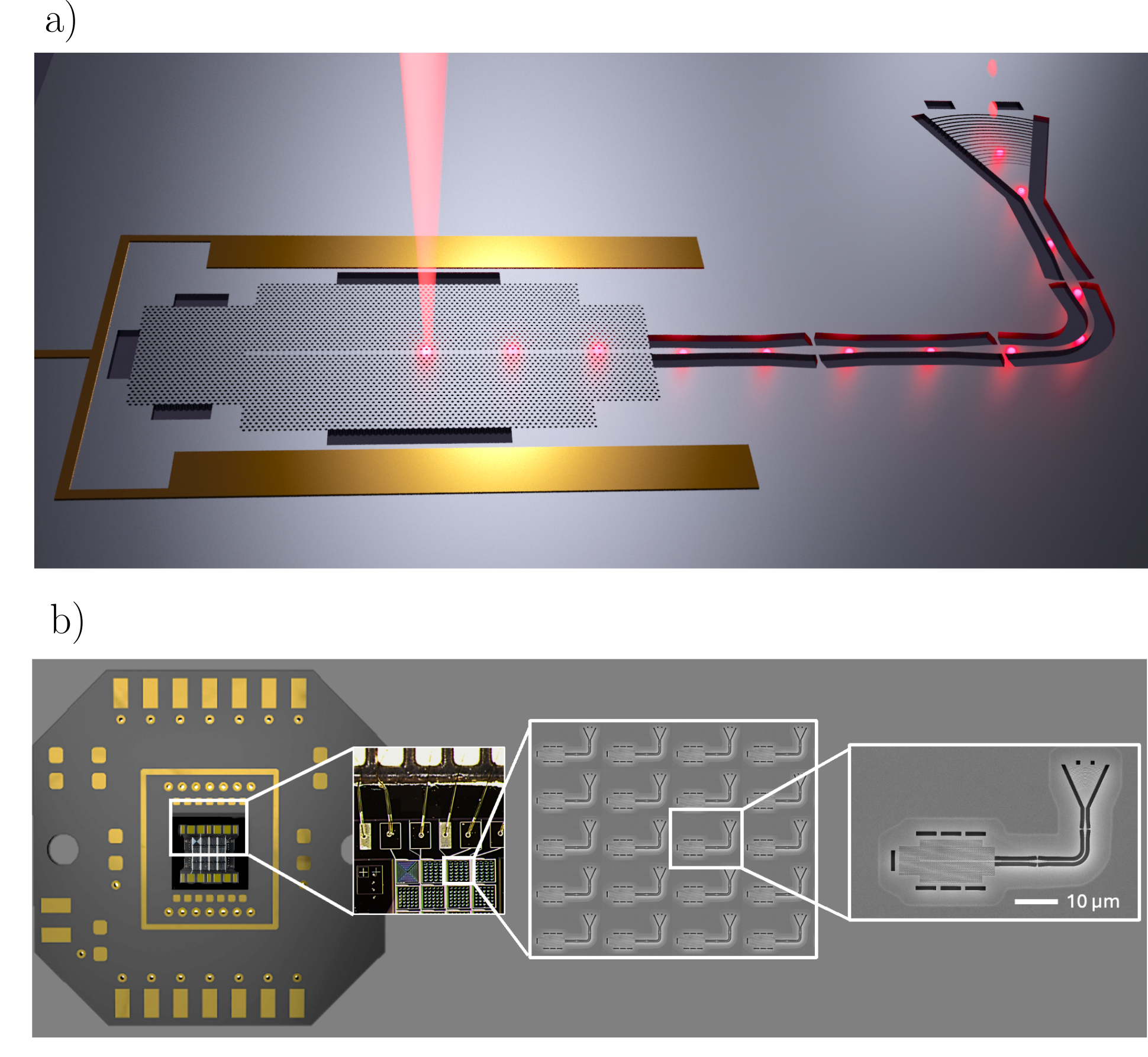}
\caption{\textbf{Illustration of operational principle and chip layout of planar QD single-photon sources}.
a) Illustration of a QD single-photon source based on a planar photonic crystal waveguide. A pulsed laser repetitively excites the QD that subsequently emits single photon pulses into a photonic crystal waveguide that routes the train of photons to an outcoupling grating for high-efficiency coupling to a single-mode optical fibre. b) Image of single-photon source chip mounted on a printed circuit board (PCB). The external connectors on the PCB allow electrical access to the chip, enabling spectral tuning and the suppression of electrical noise. Insets: (i) Optical microscope image showing the overall chip layout, which consists of multiple blocks of nanostructures arranged in a grid. (ii) SEM image of a representative block containing an array of individual photonic nanostructures, allowing operation of multiple QD sources. (iii) High-magnification SEM view of a single photonic nanostructure.
\label{fig: sparrow_waveguide_all}}
\end{figure}

In the following, we will give a brief account of the operational principle of the planar photonic crystal waveguide single-photon sources. For an in-depth account, we refer the readers to Ref.~\cite{uppuscalable2020}. Figure \ref{fig: sparrow_waveguide_all}a) illustrates the basic physical principles. A QD is illuminated by short laser pulses, exciting an electron from the valence band to the conduction band to form an electron-hole pair (an exciton). Discrete QD energy levels are ensured by using composite semiconductor materials, resulting in a ``particle-in-a-box" quantum confinement potential. The exciton subsequently recombines by spontaneous emission, emitting a single photon that is collected by the photonic nanostructure encompassing the QD. See Ref.~\cite{lodahlInterfacingSinglePhotons2015} for an in-depth description of QDs and photonic nanostructures. 

\begin{figure}[htb!]
\centering
\includegraphics[width=.9\linewidth]{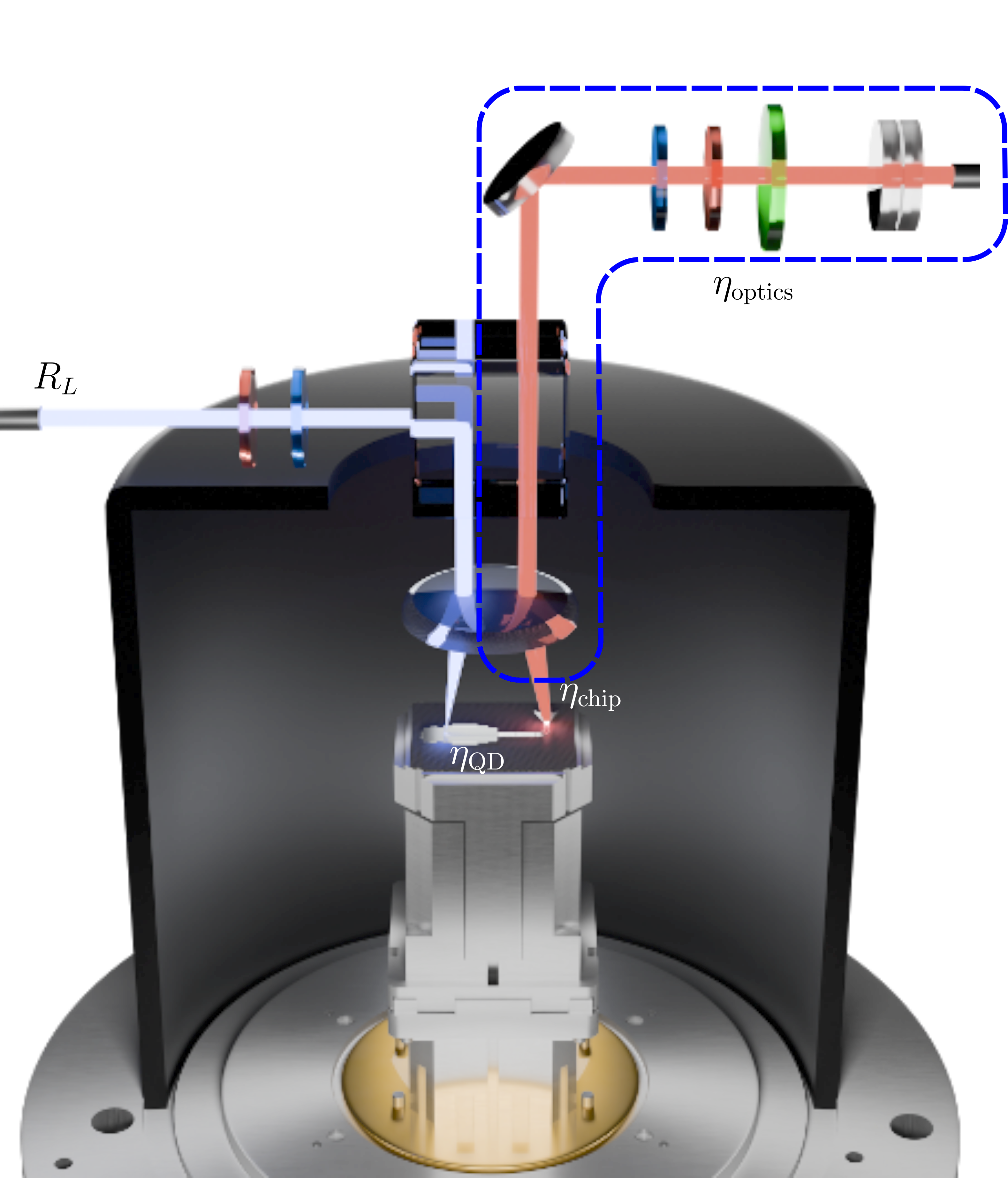}
\caption{\textbf{Full experimental setup for extracting single photons from the chip}. A pulsed laser, with repetition rate $R_\mathrm{L}$ is sent via a beamsplitter into the cryostat. The polarisation is controlled by quarter- and half-wave plates after which the light is focused by an objective lens onto the single-photon chip, which is cooled to $\sim 4$K. 
The internal efficiency of the QD emitter is $\eta_\mathrm{QD}$, which specifies the probability of emitting a photon within the zero-phonon line and into the waveguide after being excited by the laser pulse. The collected photon subsequently propagates from the QD, through the chip and into free space with an efficiency $\eta_\mathrm{chip}$. 
The out-coupled light is sent upward, collected by the same lens and transmitted by the beamsplitter thereby separating the single-photon emission from the input light. Here, it encounters the first lens of the free space optics, where the brightness is measured. $\eta_\mathrm{optics}$ is the total transmission efficiency of the free-space optics, which controls photon polarisation, implements phonon sideband filtering, and finally couples the single photons into an optical fibre. The complete setup is fully automatized and can therefore be operated ``hands free" constituting a full plug-and-play single-photon source system. 
\label{fig: 3D_setup_full}}
\end{figure}

Photonic nanostructures can tailor the optical density of states, which is the physical principle used to control single-photon emission. One approach is to use a high Q-factor cavity to Purcell enhance the emission. However, the working bandwidth with this approach is limited (typically on the order of $\sim0.1$~nm), 
as it is determined by the inverse of the cavity Q-factor. 
In contrast, a photonic crystal waveguide exploits strong suppression of spontaneous emission by a photonic band gap as the primary mechanism for enhancing light-matter coupling, offering a broad spectral bandwidth (up to $\sim 10$~nm) and has a high tolerance to the QD's position in the waveguide~\cite{lodahlInterfacingSinglePhotons2015}. On top of this suppression, broadband Purcell enhancement into the slow-light waveguide mode yields record high internal coupling efficiencies ($\beta > 98.4 \%$)~\cite{arcariNearUnityCouplingEfficiency2014}, with even higher efficiencies readily within reach~\cite{lodahlInterfacingSinglePhotons2015}. A near-unity $\beta$-factor is essential for realising a high photon-emitter quantum cooperativity as required for advanced applications in quantum gates and deterministic entanglement generation \cite{uppuQuantumdotbasedDeterministicPhoton2021}. For concrete comparison, a value of $\beta=99\%$ which is readily achievable with QD waveguide devices would, in a cavity platform, require a cavity Purcell factor of $\sim 100$---far beyond currently demonstrated capabilities with QD cavity sources.
Of more practical importance, waveguide-based single-photon sources are highly robust, maintaining their high-performance through multiple thermal cycles. In contrast, cavities are highly spectrally-sensitive to nearby static charge configurations, which can shift operational wavelengths after temperature cycles.
Finally, in a planar waveguide platform, the pump laser is applied at a point spatially distinct from the collection point, enabling strict resonant excitation with high laser extinction without reducing the collected single-photon signal. For all these reasons, QDs in planar waveguides have emerged as a premier platform for scalable single-photon sources.

The layout of the photonic chip comprising photonic crystal waveguide (PCW) single-photon sources is shown in  Fig.~\ref{fig: sparrow_waveguide_all}b. Electrical access to the chip allows external voltage control of the QDs, allowing spectral tuning and reduction of electrical noise to achieve near transform-limited optical linewidths~\cite{pedersenTransformLimitedQuantumDot2020}.
An additional advantage of planar waveguide devices is that a tunnel-coupled back gate can stabilise and control the QD charge state. This is difficult to implement in micropillar cavities, where the required doped layers would sit near field maxima and introduce prohibitive free-carrier absorption~\cite{najerSuppressionSurfaceRelatedLoss2021}, making charge-state locking and hence suppression of blinking much harder.
The device contains a grid of hundreds of PCWs, see Fig.~\ref{fig: sparrow_waveguide_all}c, each containing several efficiently coupled QDs. The waveguide collects the emitted photons into a single mode, and the photons are subsequently routed out of the PCW section through nanobeam waveguides and to a specifically designed grating outcoupler tailored to efficient fibre coupling, see Fig.~\ref{fig: sparrow_waveguide_all}a.

\begin{figure*}[htb!]
\centering
\includegraphics[width=\linewidth]{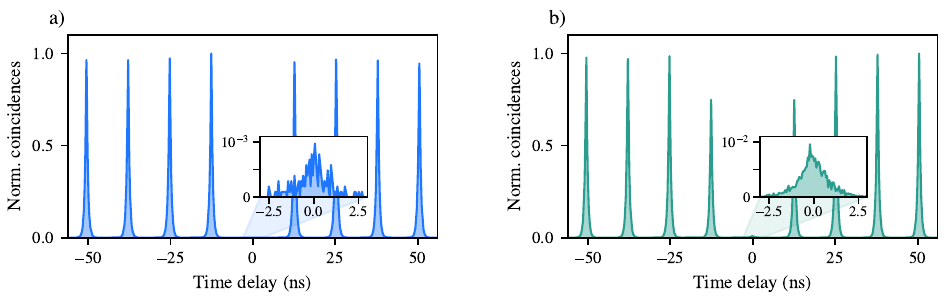}
\caption{\textbf{Single photon purity and Hong-Ou-Mandel (HOM) visibility of state-of-the-art planar single-photon sources}. a) Measurement of second order autocorrelation function $g^{(2)}(\tau)$ versus time delay, where the suppression of coincidences at zero time delay signals single-photon emission. The purity is quantified by the minor residual coincidences detected at zero time delay (see inset) and we find $g^{(2)}(0){=}\left(0.1{\pm}0.1\right)\%$. b) Histogram of coincidences versus time delays resulting from a HOM experiment where subsequently emitted photons from the QD are interfered on a symmetric beamsplitter. The indistinguishability is quantified by the minor residual coincidences detected at zero time delay (see inset) and we find $V = \left(97.1\pm0.1\right)\%$. Note that this constitutes an experimental lower bound with no corrections for multi-photon contributions or measurement setup asymmetries.   
\label{fig: g2_plot}}
\end{figure*}

Obtaining the highest performances single-photon sources requires strict resonant excitation of the QD transition. Indeed, non-resonant excitation can introduce additional phonon broadening that limits photon coherence and can also induce optical blinking, reducing the overall source efficiency. Because phonon broadening is strongly reduced at cryogenic temperatures~\cite{tighineanuPhononDecoherenceQuantum2018}, the standard operation temperature of the QD single-photon sources is 4~K. Inelastic phonon scattering leads to broad phonon sidebands, typically containing $5-10$~\% of the emitted intensity at 4~K for QDs in bulk samples~\cite{lodahlInterfacingSinglePhotons2015}. This contribution can be further reduced by Purcell enhancement of the QD zero-phonon  line---either in a cavity or in a waveguide---and because the remaining phonon sideband emission is broadband, it can be readily removed with spectral filters. Elastic phonon scattering broadens the QD zero-phonon line and imposes the fundamental limit on photon coherence (i.e., indistinguishability). A route to reducing elastic phonon scattering processes has also been identified~\cite{Dreeßen_2019}, providing a roadmap to sources with near-unity indistinguishability. 

The practical operation of the single-photon source requires interfacing with optical beams, i.e., a classical laser for optical excitation is coupled in through one port and the non-classical signal is collected through another. Fig.~\ref{fig: 3D_setup_full} depicts such an optical apparatus: a pulsed laser with repetition rate $R_\text{L}$ is directed along an excitation path to resonantly excite the QD, while the emitted single photons are collected along a separate path from a grating coupler on the chip into a single-mode fibre. Polarisation control on both the excitation and collection paths ensures that a single neutral-exciton dipole is excited, and that the collected signal is aligned to the grating coupler polarisation. The key experimental control parameters required to operate the source include spatial alignment, polarization, wavelength, pulse duration, and optical power of the excitation laser, as well as the applied DC bias voltage across the chip. These parameters can be automatically controlled and tracked, thereby offering a fully automated plug-and-play single-photon source product~\cite{sparrownest}. The performance and relevant figures-of-merit of these sources are described in the following section.

\section{Figures of merit of deterministic single-photon sources}

Three main figures-of-merit benchmark the performance of a deterministic single-photon source: i) source purity, ii) indistinguishability of emitted photons, and iii) overall source efficiency. In the following we describe each of the parameters, detail how to obtain them experimentally, and show examples of state-of-the-art data realised with PCW sources. 

\subsection{Single-photon source purity}

The single-photon source purity relates to the photon-number occupation in each emitted pulse, and quantifies how close the source is to producing one, and only one, photon upon excitation. Therefore, a purity of $100\%$ indicates that the source is emitting optical pulses containing single-photon Fock states $|1\rangle$. The second-order autocorrelation function at zero delay is $g^{(2)}(0){=}{\langle \hat{n}(\hat{n}-1)\rangle}/{\langle \hat{n}\rangle^2}$, where $\hat{n}$ is the number operator. This autocorrelation vanishes for a single-photon state and the single-photon purity is defined as $\mathcal{P}{=}1{-}g^{(2)}(0)$, which is valid in the anti-bunched regime where $ 0 \leq g^{(2)}(0) < 1$. 

A Hanbury Brown and Twiss (HBT) experiment can be used to directly measure the single-photon purity. 
Here photons from the single-photon source are directed to a beamsplitter (of any reflectivity), where a detector at each output records counts as a function of time. By measuring coincidences between the two detectors, photon correlations are recorded and used to reconstruct the second order correlation function $g^{(2)}(\tau)$, where $\tau$ is the time difference between detector clicks. Fig.~\ref{fig: g2_plot}a presents examples of such data. For an ideal single-photon source, no simultaneous coincidences can occur, and therefore the absence of coincidences at zero time delay in an HBT measurement demonstrates that detection events arise from only single photons. 

Experimentally,  $g^{(2)}(0)$ is obtained as
\begin{equation}
    g^{(2)}(0) = \frac{C_0}{C},
\end{equation}

where $C_0$ is the integrated area of the central (zero time delay) peak within a chosen time window, and $C$ is the mean integrated area of the uncorrelated peaks arising from subsequently emitted photons in the pulsed HBT measurement. Fig~\ref{fig: g2_plot}a shows examples of HBT data of a PCW single-photon source operated under $\pi$-pulse excitation of a neutral exciton. Using a 4~ns integration window (much longer than the photon lifetime), we obtain $g^{(2)}(0){=}\left(0.1{\pm}0.1\right)\%$, hence a near-unity purity of $\mathcal{P}{=}\left(99.9{\pm}0.1\right)\%$. 

In order to achieve such high single-photon purity, several sources of potential error must be carefully addressed. The excitation pulse duration should be considerably shorter than the QD decay time to strongly suppress re-excitation processes~\cite{Dada:16}, and the excitation laser leakage into the PCW must also be minimised. The present data were taken  using ${\sim}10~$ps long laser pulses, compared to values of 250-410~ps for the QD lifetimes in this work. Furthermore, resonant excitation of the QD through weak leaky waveguide modes entail only weak scattering of laser light into the waveguide. The light leaked into the waveguide may even destructively interfere with light emitted from the QD (depending on the exact QD position in the PCW) to produce even higher single-photon source purity~\cite{gonzalez-ruizTwophotonCorrelationsHongOuMandel2025}.

\subsection{Photon-photon indistinguishability}

Photon indistinguishability describes the extent to which subsequently emitted photons from the QD are identical to each other, i.e., having the same frequency, polarisation, and temporal and spatial profiles. 

 The experimental method of measuring photon indistinguishability applies a Hong-Ou-Mandel (HOM) interferometer, where two input photons impinge simultaneously on a balanced (a $50\%/50\%$ transmission/reflection) beamsplitter, and exit it together via either output as a bunched pair. The degree of indistinguishability is determined by the quantum interference visibility of the HOM interferometer, i.e., the degree to which the two impinging photons can destructively interfere relative to two fully distinguishable photons.

\begin{figure}[htb!]
\centering
\includegraphics[width=\linewidth]{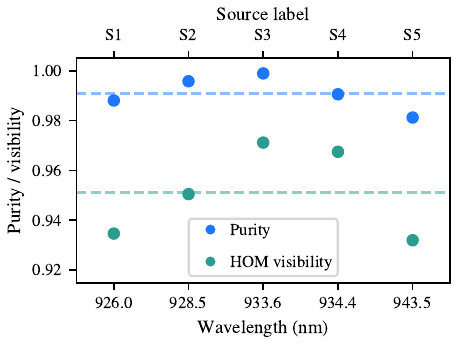}
\caption{\textbf{Data of high-performance single-photon sources.} Multiple sources are characterised under $\pi$-pulse excitation, yielding an average single-photon purity (dashed blue line) exceeding $99\%$, and an average raw HOM visibility (dashed green line) above $95\%$. Source $\mathrm{S}3$ corresponds to the maximum values quoted in the main text. 
\label{fig: g2_hom-sources}}
\end{figure}

To experimentally extract the HOM visibility, the QD single-photon sources are repeatedly operated at a pulsed repetition rate of 80~MHz, and the emitted photons directed to a HOM interferometer---i.e., a Mach-Zehnder interferometer with one arm delayed corresponding to the laser repetition rate. Two subsequently emitted photons may simultaneously arrive and destructively interfere on the output HOM beamsplitter, which is monitored by coincidence single-photon detection between the two output ports.  The HOM visibility of such an experiment is determined by
\begin{equation}
    V = 1- 2\frac{C_0}{C},
    \label{eq:HOM}
\end{equation}
 where $C_0$ is the integrated area of a central peak at zero time delay, and $C$ is the mean integrated area of the uncorrelated peaks (the correlation peaks corresponding to time delays of at least two laser repetition periods). Examples of HOM correlation data are shown in Fig~\ref{fig: g2_plot}b. We note that fully distinguishable single photons, ($V{=}0$), bunch $50\%$ of the time, i.e., $C_0 = C/2$, and fully indistinguishable single photons, $V{=}1$, bunch $100\%$ of the time, i.e., $C_0 = 0$.

\begin{figure*}[t!]
\centering
\includegraphics[width=\linewidth]{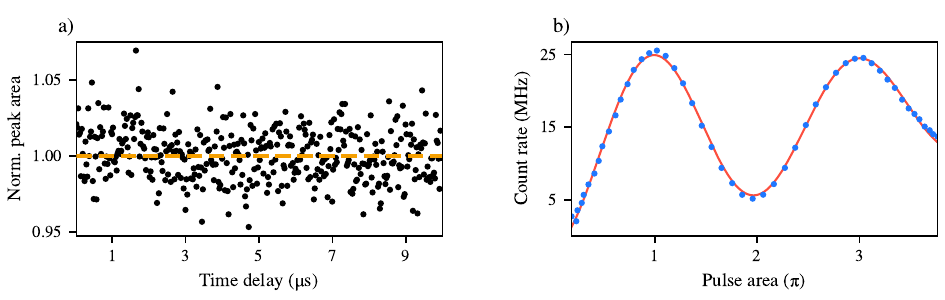}
\caption{\textbf{Demonstration of blinking free high-efficiency single-photon source.} a) Long time-scale autocorrelation measurement. The data is binned with a 25~ns bin width, corresponding to twice the repetition rate of the laser, avoiding aliasing effects from data sampling. Since the peak area remains equal throughout the 10$\mu s$ collection run, this measurement shows an absence of blinking processes, which would manifest as a decay of the peak area. b) Observation of Rabi oscillations in resonant excitation of QD devices. Directly measured (blue points) single photon count rates versus excitation power resulting from the coherent driving of source $\mathrm{S}1$. The detected count rate at $\pi$-pulse excitation here is $25.5$~MHz, slightly differing from the maximum value reported in the main text due to a different detection efficiency condition. The red curve is a fit to the data.  
\label{fig:Rabi}}
\end{figure*}

Photon indistinguishability is a key metric in photon quantum-information processing, being a direct metric of the photonic qubit quality. In the literature on QD single-photon sources, there appears to be a rather diverse range of indistinguishability data, with various data post-processing methods used to correct for different experimental contributions (e.g., background photons, multi-photon contributions, imperfect measurement setup, etc) that are argued to be possible to correct in more idealised experimental conditions. Important caveats in these corrections have recently been identified~\cite{gonzalez-ruizTwophotonCorrelationsHongOuMandel2025}, e.g., correcting HOM data for a finite single-photon purity can be misleading and often results in artificially increased HOM values. Furthermore, in real-world applications any imperfections are important, and consequently in the present work we report exclusively ``raw HOM" data that are directly extracted from experimental data and not corrected for any imperfections, not even the slight imbalance of the HOM beamsplitter. Consequently, the reported data can be considered an experimental lower bound of attainable values. 

 Fig~\ref{fig: g2_plot}b shows representative HOM data of a PCW QD single-photon source operated at $\pi$-pulse excitation. We record a raw HOM visibility of $V{=}\left(97.1{\pm}0.1\right){\%}$ by using Eq.~\ref{eq:HOM} with a time-window of 4 ns. To our knowledge, this is the highest HOM visibility reported to date for a QD single-photon source, and proves that all additional noise processes have been reduced to such a degree that the source is limited by the fundamental phonon decoherence processes described previously. This particular QD source  simultaneously features $\mathcal{P}{=}\left(99.9{\pm}0.1\right)\%$ and $V{=}\left(97.1{\pm}0.1\right){\%}$, constituting record performance for QD sources (see further discussion below).  However, to investigate the reproducibility of the approach we additionally characterised a number of sources. Fig.~\ref{fig: g2_hom-sources} displays data for five high-performance sources recorded across different fabricated chips. Here we find average values of $\mathcal{P}>99\%$ and $V>95\%$. 

\subsection{Single-photon source efficiency}
The overall efficiency of the single-photon source is essential for the usability of the device. This is particularly important since most applications of deterministic single-photon sources rely on a de-multiplexing setup to turn one single-photon source into N identical sources for multi-photon experiments. The overall rate of multi-photon generation scales exponentially with the total photon source system efficiency, and currently $N=20$ is the record value demonstrated with a QD source~\cite{wangBosonSampling202019a}.
Several efficiency performance metrics are defined in the literature, and one quantity is the first-lens efficiency, $\eta_\text{FL} = \eta_\mathrm{QD} \eta_\mathrm{chip}$, which accounts for the actual collection and propagation in the chip (transmission $\eta_\mathrm{QD}$) and the outcoupling from the chip and into the collection lens (transmission $\eta_\mathrm{chip}$). See Fig.~\ref{fig: 3D_setup_full} for illustration of the efficiencies. 
For the end-user the more meaningful and less ambiguous metric is the overall system efficiency, $\eta_\text{S}$. The system efficiency quantifies the probability that once the QD is excited, a usable single photon (i.e. free of residual laser background, phonon sidebands, \blue{etc.}) is coupled into a single mode fibre for direct application in quantum information processing.  
$\eta_\text{S}=\eta_\text{optics}\eta_\text{FL}$, where $\eta_\text{optics}$ describes the efficiency of the optical setup used to clean and collect the emitted single photon signal (see Fig.~\ref{fig: 3D_setup_full}).

Achieving a high system efficiency requires all components in the system operating with high efficiency simultaneously: the QD photon source, the collection waveguide, the out-coupling grating, and the external filtering setup, cf. Fig.~\ref{fig: 3D_setup_full}. In particular, full resonant $\pi$-pulse inversion of the QD population is required, and the tendency of solid-state emitters to ``blink" due to coupling to non-radiative states would need to be fully overcome. Fig.~\ref{fig:Rabi} shows experimental data of such operations of a QD PCW single-photon source demonstrating $>$25 MHz detected single photons and blinking-free operations up to 10 $\mu$s (from long-time HBT data).  In the presented devices, the voltage applied across the QD mitigates blinking effects and also slow spectral diffusion due to charge noise\blue{,} see Fig.~\ref{fig:Rabi}a. 
The overall system efficiency can be measured directly with a calibrated single-photon detector. We have
\begin{equation}
    \eta_\text{S}=\frac{R_\text{s}}{R_\text{L}\eta_\text{d}},
    \label{eq:Etafibre}
\end{equation}
where $\eta_\text{d}$ is the efficiency of the superconducting nanowire single-photon detector (SNSPD) used in the experiment, and $R_s$ is the measured single photon rate.
Sources $\mathrm{S}2$ to $\mathrm{S}5$ presented in the current work  correspond to a generation of devices with system efficiencies in the range of $\eta_\text{S}=\left[20\%{-}35\%\right]$. The overall efficiency of the collection setup including all optics, spectral filters, and coupling into a single-mode fibre was $\eta_\text{optics}{=}70\%$. Source $\mathrm{S}1$ corresponds to a novel generation of devices where chip parameters, such as on-chip propagation losses and grating out-coupling efficiencies, have been optimised even further. In fact, with source S1 we directly measured $R_\text{s}=26.2$~MHz of single photons at $\pi$-pulse excitation, using a pump rate of $R_\text{L}=80$~MHz and a detection channel with efficiency $\eta_\text{d}=59.2\%$. This corresponds to a single photon source system efficiency as high as $\eta_\text{S}=55.3\%$. We note that this high efficiency is met simultaneously with $\mathcal{P}=98.8\%$ and $V=93.5\%$, as reported in Fig.~\ref{fig: g2_hom-sources}. By adding an additional spectral filter to the collection setup (to reduce multiphoton contributions), the same source can be operated (labelled $\mathrm{S}1^*$) with $\eta_\text{S}=50.5\%$, $\mathcal{P}=99.3\%$, and $V=95.0\%$. 

\begin{figure}[hbt!]
\centering
\includegraphics[width=\linewidth]{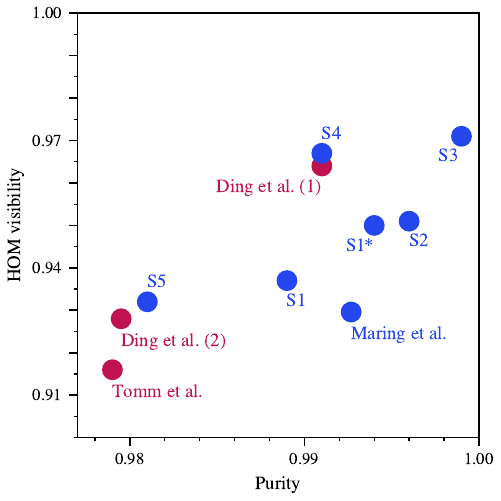}
\caption{\textbf{Comparison of QD single-photon sources based on cavities and waveguides.}. The upper right corner, with unity purity and HOM visibility, represents the ideal single photon source. The quality of PCW single-photon sources (S1-S5) are displayed alongside state-of-the-art QD single-photon sources reported in the literature. $\mathrm{S}1^*$ represents source $\mathrm{S}1$ with an additional spectral filter. 
Blue points represent commercial sources, whilst red points represent academic ones. 
\label{fig: benchmarking_plot}}
\end{figure}

We finally benchmark the reported source performances against current state-of-the-art in the literature, including both commercially available platforms (vertical micropillar cavities) and academically pursued systems (open cavities and micropillar cavities). Table~\ref{tab: efficiency_benchmark} displays the system efficiency data. Furthermore, Fig.~\ref{fig: benchmarking_plot} compares their quality, in terms of purity and indistinguishability. We highlight that two of the sources, $\mathrm{S}2$ and $\mathrm{S}3$, display the highest (uncorrected) photon purities and $\mathrm{S}3$ and $\mathrm{S}4$, the highest (uncorrected) degree of indistinguishability reported with QD sources. 

\begin{table}[htb!]
\centering
\begin{tabular}{ |c|c|c|c| }
\hline
Academic & $\eta_\mathrm{S}$ & Commercial & $\eta_\mathrm{S}$ \\ 
 \hline 
 Ding et al. (1)~\cite{dingHighefficiencySinglephotonSource2025} & 71.2\% & Sparrow $\mathrm{S}1$  & $55.3$\% \\
 Tomm et al.~\cite{tommBrightFastSource2021} &  57\% & Sparrow $\mathrm{S}1^*$  & $50.5$\% \\
 Wang et al.~\cite{wangBosonSampling202019} & 26.2\% & Maring et al.~\cite{maringVersatileSinglephotonbasedQuantum2024} &  28.9\%\footnote{Using the quoted 92\% detection efficiency} \\
 Ding et al. (2)~\cite{dingOnDemandSinglePhotons2016} &  14\%\footnote{Using the quoted $\sim 33\%$ detection efficiency.} & & \\ 
 \hline
\end{tabular}
\caption{\textbf{Photon source system efficiencies.} The highest reported values for QD based single-photon sources. Left, results from academic research groups. Right, commercially available photon sources.}
\label{tab: efficiency_benchmark}
\end{table}

~
\\
In conclusion, we have presented experimental data on photonic crystal waveguide based single-photon sources including demonstrated state-of-the-art performance on purity, indistinguishability, and system efficiency. We conclude that PCW deterministic single-photon source performance compares favourably to cavity based platforms, primarily due to the advantages of high internal coupling efficiency and the ease of strict resonant excitation. Furthermore, the robustness and ease of operation may become a major practical advantage when scaling up this technology in the coming years. Nonetheless, the already demonstrated sources can readily be applied in multi-photon experiments by implementing de-multiplexing using electro-optic switching and fibre delay loops, and further device optimizations and the prospects of O-band operation entail that even better device performance will be continuously demonstrated. The extension of these sources to realise deterministic multi-photon entanglement sources promise to enable even more powerful quantum photonic resources. Quantum dot light-matter interfaces have reached the level of maturity that serious technology development is starting to take place. The ultimate ``push-of-a-button" generators of advanced quantum states of light---the main missing link for deployable photonic quantum information processing---now appear to be within reach. 

\bibliography{SparrowBib}
\end{document}